\documentclass[aps,pra,superscriptaddress,twocolumn,showpacs,floatfix]{revtex4-1}

\usepackage{amsmath}
\usepackage{amssymb}
\usepackage{bm}
\usepackage{graphicx}
\usepackage{appendix}
\usepackage{longtable}
\usepackage{float}
\usepackage{arydshln}
\usepackage{color}
\begin{document}
\title{High reflectance and optical dynamics of a metasurface\\ comprizing quantum $\Lambda$-emitters}

\author{Igor V. Ryzhov}
\affiliation{Herzen State Pedagogical University, St. Petersburg, 191186 Russia}

\author{Ramil F. Malikov}
\affiliation{M. Akmullah Bashkir State Pedagogical University, 450008 Ufa, Russia}%

\author{Victor\ A.\ Malyshev}
\affiliation{Zernike Institute for Advanced Materials, University of Groningen, Nijenborgh 4, 9747
AG Groningen, The Netherlands}
\affiliation{Herzen State Pedagogical University, St. Petersburg, 191186 Russia}

\date{\today}

\begin{abstract}

In this Letter, we study theoretically reflectance of a monolayer comprizing regularly spaced quantum $\Lambda$-emitters. Due to high density of the latter, the monolayer almost totally reflects the incident field in the vicinity of the system's collective (excitonic) resonance. The emitter self-action through the secondary field provides a positive feedback, interplay of which with the inherent nonlinearity of an emitter itself, results in an exotic behavior of the system reflectance, including bistability, self-oscillations, and chaotic dynamics. All these features might be of interest for nanophotonic applications.

\end{abstract}

\maketitle

{\bf Introduction}.
Nowdays, (meta)surfaces composed of meta-atoms have received a greate deal of attention due to their exceptional abilities in light manipulation and versatility in sub-wavelength nanophotonics applications~\cite{ChenRepProgrPhys2016,HsiaoSmallMethods2017,ChangAnnuRevMaterRes2018}.
Recently, it has been reported that an atomically thin layer of MoSe$_2$ encapsulated by hexagonal boron nitride manifests high reflectance in the vicinity of collective (excitonic) resonance~\cite{Back2018,Scuri2018}. Likewise, quantum metasurfaces of arrays of atoms trapped in an optical lattice~\cite{BekensteinNatPhys2020}
and two-dimensional supercrystals of semiconductor quantum dots (SQDs)~\cite{Evers2013,BaimuratovSciRep2013,BaimuratovOptLett2017} exhibit similar behavior~\cite{RyzhovPRA2019,BayramdurdiyevJETP2020}. Moreover, the optical response of the latter, in addition, may demonstrate multistability and instabilities of different types, such as periodic and aperiodic self-oscillations and dynamical chaos.

Here, we are modelling reflection of quasi-resonant radiation from a monolayer of quantum emitters with the $\Lambda$ arrangement of energy levels. Doped quantum dots~\cite{Brunner2009} and organic nanocrystals with vibronic structure of the ground state~\cite{BookNanocrystal2011} can be considered as examples of such a type of emitters. The (secondary) field acting on a given emitter on the part of the others is taken into account. This field provides an intrinsic positive feadback, interplay of which with nonlinearity of the emitters themselves gives rise to instabilities of the monolayer reflectance. Similarly to supercrystals of ladder-~\cite{RyzhovPRA2019} and V-type~\cite{BayramdurdiyevJETP2020} emitters, we found bistability, periodic and aperiodic self-oscillations, and chaotic behavior of  reflectance. All these properties are demanding for nanophotonics.

{\bf Model and formalism}.
Our model system consists of a $N\times N$ square lattice of identical quantum emitters having a single upper state $|3 \rangle$ and a doublet $|1 \rangle$ and $|2 \rangle$ in the lower state. Optical transitions are allowed only between the upper state $|3\rangle$ and those of the doublet $|1\rangle$ and $|2\rangle$ (so called $\Lambda$-emitter). These transitions are characterized by the transition dipole moments $\bf{d}_{31}$ and $\bf{d}_{32}$ which, for the sake of simplicity, are set to be real and parallel to each other, so that ${\bf d}_{32} = \mu\bf{d}_{31}$. The upper state $|3\rangle$ decays spontaneously to the states of the doublet $|2\rangle$ and $|1\rangle$ with rates $\gamma_{31}$ and $\gamma_{32} = \mu^2 \gamma_{31}$, respectively. The doublet splitting $\Delta_{21}$ is assumed to be small compared to the optical transition frequencies $\omega_{31}$ and $\omega_{32}$.
Relaxation within the doublet is accounted for by a constant $\gamma_{21}$. The monolayer undergoes a quasi-resonant continuous wavw (CW) external field $\bm{\mathcal E} = {\bf E}_0 \cos(\omega_0 t)$ of amplitude ${\bf E}_0$ and frequency $\omega_0$ incident normally to the monolayer and polarized along the transition dipole moments.

Optical dynamics of a given $\Lambda$-emitter in the monolayer is governed by the system of equations for the density matrix $\rho_{\alpha\beta}$ ($\alpha,\beta$ = 1,2,3), which within the mean-field and rotating wave approximation reads
\begin{subequations}
\label{allrho}
\begin{equation}
\label{rho11}
    \dot{\rho}_{11} = \gamma_{21} \rho_{22} + \gamma_{31} \rho_{33} + \Omega^* \rho_{31} + \Omega \rho_{31}^*~,
\end{equation}
\begin{equation}
\label{rho22}
    \dot{\rho}_{22} = -\gamma_{21} \rho_{22} + \gamma_{32} \rho_{33}  + \mu(\Omega^* \rho_{32} + \Omega\rho_{32}^*)~,
\end{equation}
\begin{eqnarray}
\label{rho33}
    \dot{\rho}_{33} = &-& (\gamma_{31} + \gamma_{32}) \rho_{33} - \Omega^* \rho_{31} - \Omega \rho_{31}^*
    \nonumber\\
    &-& \mu (\Omega^* \rho_{32} + \Omega \rho_{32}^*)~,
\end{eqnarray}
\begin{eqnarray}
\label{rho31}
    \dot{\rho}_{31} = &-& \left[ i\Delta_{31} + (\gamma_{31} + \gamma_{32})/2 \right] \rho_{31}
    \nonumber\\
    &+& \Omega(\rho_{33} - \rho_{11}) - \mu \Omega \rho_{21}~,
\end{eqnarray}
\begin{eqnarray}
\label{rho32}
    \dot{\rho}_{32} = &-& \left[ i\Delta_{32} + (\gamma_{31} + \gamma_{32} + \gamma_{21})/2 \right] \rho_{32}
    \nonumber\\
    &+& \mu\Omega(\rho_{33} - \rho_{22}) - \Omega \rho_{21}^*~,
\end{eqnarray}
\begin{equation}
\label{rho21}
    \dot{\rho}_{21} = - \left( i\Delta_{21} + \gamma_{21}/2  \right) \rho_{21} + \mu \Omega^* \rho_{31} + \Omega \rho_{32}^*~.
\end{equation}
\begin{equation}
\label{Local field}
    \Omega = \Omega_0 + (\gamma_R - i\Delta_L)(\rho_{31} + \mu\rho_{32})~,
\end{equation}
\end{subequations}
where $\Delta_{31} = \omega_0 - \omega_{31}$ and $\Delta_{32} = \omega_0 - \omega_{32}$ are detunings of the incident field frequency $\omega_0$ away from the resonance frequences $\omega_{31}$ and $\omega_{32}$ of the $1 \leftrightarrow 3$ and $2 \leftrightarrow 3$ transitions, respectively. Furthermore, $\Omega = d_{31}E/\hbar$, given by Eq.~\ref{Local field}, is the Rabi amplitude of the mean field with $E$ being the amplitude of the latter, $\hbar$ is the reduced Plank constant, $\Omega_0 = d_{31}E_0/\hbar$ stands for the Rabi amplitude of the incident field. The second term in Eq.~\ref{Local field} represents the Rabi amplitude of the secondary field produced by all other emitters at the position of a given one. A part proportional to $\gamma_R$ describes the far-zone contribution to the secondary field, while th one scaled with $\Delta_L$ accounts for the near-zone part which is analogous to the Lorentz local field~\cite{Benedict1991}. The constants $\gamma_R$ and $\Delta_L$ are given by ~\cite{RyzhovPRA2019}
\begin{subequations}
\begin{equation}
\label{gammaRpoint-like}
    \gamma_R = (3/8)\gamma_{31} N^2~, \quad Na \ll \lambda^\prime~,
\end{equation}
\begin{equation}
\label{gammaRextended}
    \gamma_R = 4.5 \gamma_{31} (\lambda^\prime/a)^2~, \quad Na \gg \lambda^\prime~,
\end{equation}
\begin{equation}
\label{DeltaLextended}
    \Delta_L  = 3.4 \gamma_{31} (\lambda^\prime/a)^3~,
\end{equation}
\end{subequations}
where $\lambda^\prime = \lambda/(2\pi)$ is the reduced wavelength.
As follows from Eq.~(\ref{gammaRpoint-like}), for the point-like system ($\lambda^\prime \ll Na$) $\gamma_R$ is determined by the total number of emitters in the system, $N^2$, while in the case of an extended sample [$\lambda^\prime \gg Na$, Eq.~(\ref{gammaRextended}], $\gamma_R$ is proportional to the number of emitters within the area of ${\lambda^\prime}^2$: those emitters radiate in phase, and $\gamma_R$ is the Dicke's superradiant constant~\cite{Dicke1954,BenedictBook1996,RyzhovPRA2019} accounting for the collective radiation relaxation of $\Lambda$-emitters in the monolayer.

The parameter $\Delta_L$ is almost independent of the system size; it is nothing but the near-zone dipole-dipole interaction of a given $\Lambda$-emitter with all others. It determines the (excitonic) energy level renormalization~\cite{Benedict1991,RyzhovPRA2019,RyzhovArxiv2020} (see below). Irrespectively of the system size, $\Delta_L \gg \gamma_R$ for a dense sample ($\lambda^\prime \gg a$).

Note that Eqs.~(\ref{rho11}) -- (\ref{rho21}) conserve the total population $\rho_{11} + \rho_{22} + \rho_{33} = 1$, i.e. we consider the spontaneous decay to be the only channel of relaxation. Pure dephasing of the $\Lambda$-emitter states is neglected and will be addressed elsewhere.
We are interested in the monolayer reflectance $R$ (the reflection coefficient of light flow) which is defined as
\begin{subequations}
\begin{equation}
\label{Reflectance}
    R = \left|\frac{\Omega_\mathrm{refl}}{\Omega_0}\right|^2~,
\end{equation}
\begin{equation}
\label{Reflected field}
    \Omega_\mathrm{refl} = \gamma_R(\rho_{31} + \mu\rho_{32})~,
\end{equation}
\end{subequations}
where $\Omega_\mathrm{refl}$ is the Rabi amplitude of the reflected field~\cite{RyzhovArxiv2020}.

{\bf Results}.
In our numerical calculations we used the set of parameters adjusted to 2D supercrystals of SQDs~\cite{Evers2013} (see also Ref.~\cite{RyzhovPRA2019}): $\gamma_{31} = \gamma_{32} \approx 3\cdot 10^9$~s$^{-1}$ ($\mu = 1$). The magnitudes $\gamma_R$ and $\Delta_L$ depend on the ratio $\lambda^\prime/a$. For $\lambda^\prime \sim 100 \div 200$ nm and $a \sim 10 \div 20$ nm, $\gamma_R \sim 10^{12}$ s$^{-1}$  and $\Delta_L \sim 10^{13}$ s$^{-1}$. To be specific, we set $\gamma_R = 100\gamma_{31}$ and $\Delta_L = 1000\gamma_{31}$. In what follows, the spontaneous emission rate $\gamma_{31}$ is used as the unit of all frequency-dimensional quantities, while $\gamma_{31}^{-1}$ as the time unit.

{\it Steady-state}.
First, we address the steady-state reflectance, setting to zero all time derivatives in Eqs.~(\ref{rho11})--(\ref{rho21}), Consider Eqs.~(\ref{rho31}) and~(\ref{rho32}) for $\rho_{31}$ and $\rho_{32}$ which determine the reflectance $R$, Eqs.~(\ref{Reflectance}) and~(\ref{Reflected field}). Substituting therein Eq.~(\ref{Local field}) for the mean-field Rabi amplitude $\Omega$, we get
\begin{subequations}
\begin{eqnarray}
\label{R31 extended}
    \left[ i\Delta_{31} + \Gamma_{31}
    - (\gamma_R - i\Delta_L)(Z_{31} -  \mu\rho_{21})\right] \rho_{31}
    \nonumber\\
    -\mu(\gamma_R - i\Delta_L)(Z_{31} -\mu\rho_{21}) \rho_{32}
    = \Omega_0 (Z_{31} - \mu\rho_{21})~,
\end{eqnarray}
\begin{eqnarray}
\label{R32 extended}
    \left[ i\Delta_{32} + \Gamma_{32}
    - \mu(\gamma_R - i\Delta_L)(\mu Z_{32} - \rho_{21}^*)\right] \rho_{32}
    \nonumber\\
    - (\gamma_R - i\Delta_L)(\mu Z_{32} - \rho_{21}^*) \rho_{31}
    = - \Omega_0 (\mu Z_{32} - \rho_{21}^*)~,
\end{eqnarray}
\end{subequations}
where we denoted $\Gamma_{31} = \frac{1}{2}(\gamma_{31} + \gamma_{32})$, $\Gamma_{32} = \frac{1}{2}(\gamma_{31} + \gamma_{32} + \gamma_{21})$, $Z_{31} = \rho_{33} - \rho_{11}$, and $Z_{32} = \rho_{33} - \rho_{22}$.
Equations~(\ref{R31 extended}) and~~(\ref{R32 extended}) describe two coupled nonlinear oscillators driven by two incident forces. It should be especially stresed that all characteristics of these oscillators (frequencies, relaxation rates, coupling strengths, and driving forth amplitudes) depend on the current state of the $\Lambda$-emitter. This originates direct from the secondary field acting on a given $\Lambda$-emitter on the part of the others. The consequence of this action is twofold. On one hand it results, first, in a renormalization of the transition frequencies, that is $\omega_{31} \rightarrow \omega_{31} + \Delta_L Z_{31} + \mu {\cal I}m[(\gamma_R - i\Delta_L)\rho_{21}]$ and $\omega_{32} \rightarrow \omega_{32} + \mu^2\Delta_L Z_{21} + \mu {\cal I}m[(\gamma_R - i\Delta_L)\rho_{21}^*]$ for transitions $1 \leftrightarrow 3$ and $2 \leftrightarrow 3$, respectively, and second, in an additional damping of these transitions described, accordingly, by $-\gamma_R Z_{31} + \mu {\cal R}e[(\gamma_R - i\Delta_L)\rho_{21}]$ and $-\mu^2\gamma_R Z_{32} + \mu {\cal R}e[(\gamma_R - i\Delta_L)\rho_{21}^*]$.

On the other hand, the secondary field couples the oscillators to each other [the second terms in the right-hand sides of Eqs.~(\ref{R31 extended})  and~(\ref{R32 extended})]: $\rho_{31}$ to $\rho_{32}$, with the coupling strengths $(\gamma_R - i\Delta_L)(Z_{31} - \mu\rho_{21})$, and $\rho_{32}$ to $\rho_{31}$, with the strength $(\gamma_R - i\Delta_L)(\mu Z_{32} - \rho_{21}^*)$. In the linear regime ($|\Omega_0| \ll 1$), the oscillators are decoupled, because $Z_{32} = \rho_{32} = \rho_{21} \approx 0$. However, they do couple as soon as the upper doublet state $|2\rangle$ is populated, which occurs immediately after population of the emitter higher state $|3\rangle$ and subsequent decay of the latter to the upper state $|2\rangle$ of the doublet. Interconnection of the transitions $2\leftrightarrow 1$ and $3\leftrightarrow 2$ results in an additional coupling-driven renormalization of the transition frequencies and relaxation rates. In what follows, we will refer to the above secondary-field-driven renormalization as to dressing of the $\Lambda$-emitter. We stress ones again that the overall effect of the renormalization depends on the current state of the $\Lambda$-emitter itself, which finally gives rise to a complicated behavior of the monolayer optical response as a function of the system parameters and the incident field magnitude, both in the steady state and in the time domain.

The linear regime ($|\Omega_0| \ll 1$) can be elaborated in an analytical form. In this limit, the transition  $1 \leftrightarrow 3$ mainly  contributes to $\Omega_\mathrm{refl}$. Taking in Eq.~(\ref{R31 extended}) $Z_{31} = -1$, whereas $\rho_{32} = \rho_{21} = 0$, for $\rho_{31}$ one finds
\begin {equation}
\label{eq31 linear}
    \rho_{31} = - \frac{\Omega_0}{ i(\Delta_{31} - \Delta_L) + \frac{1}{2}(\gamma_{31} + \gamma_{32}) + \gamma_R}~.
\end{equation}
Accordingly, the reflectance $R$ is given by
\begin{equation}
\label{Reflectance linear}
    R = \frac{\gamma_R^2}{(\Delta_{31} - \Delta_L)^2 + \left[\frac{1}{2}(\gamma_{31} + \gamma_{32}) + \gamma_R\right]^2}~.
\end{equation}
This expression has a maximum at $\Delta_{31} = \Delta_L$, i.e. when the frequency of the incident field, $\omega_0$, coincides with the frequency of dressed (excitonic) $1 \leftrightarrow 3$ resonance, $\omega_{31}^\prime = \omega_{31} - \Delta_L$~\cite{RyzhovArxiv2020}. The value of $R$ at this point is nearly unity because $\gamma_R \gg \gamma_{31}, \gamma_{32}$. Thus, in close vicinity to $\Delta_{31} = \Delta_L$, the system operates as {\it a perfect reflector}.

\begin{figure}[ht!]
\centering
\fbox{\includegraphics[width=0.965\linewidth]{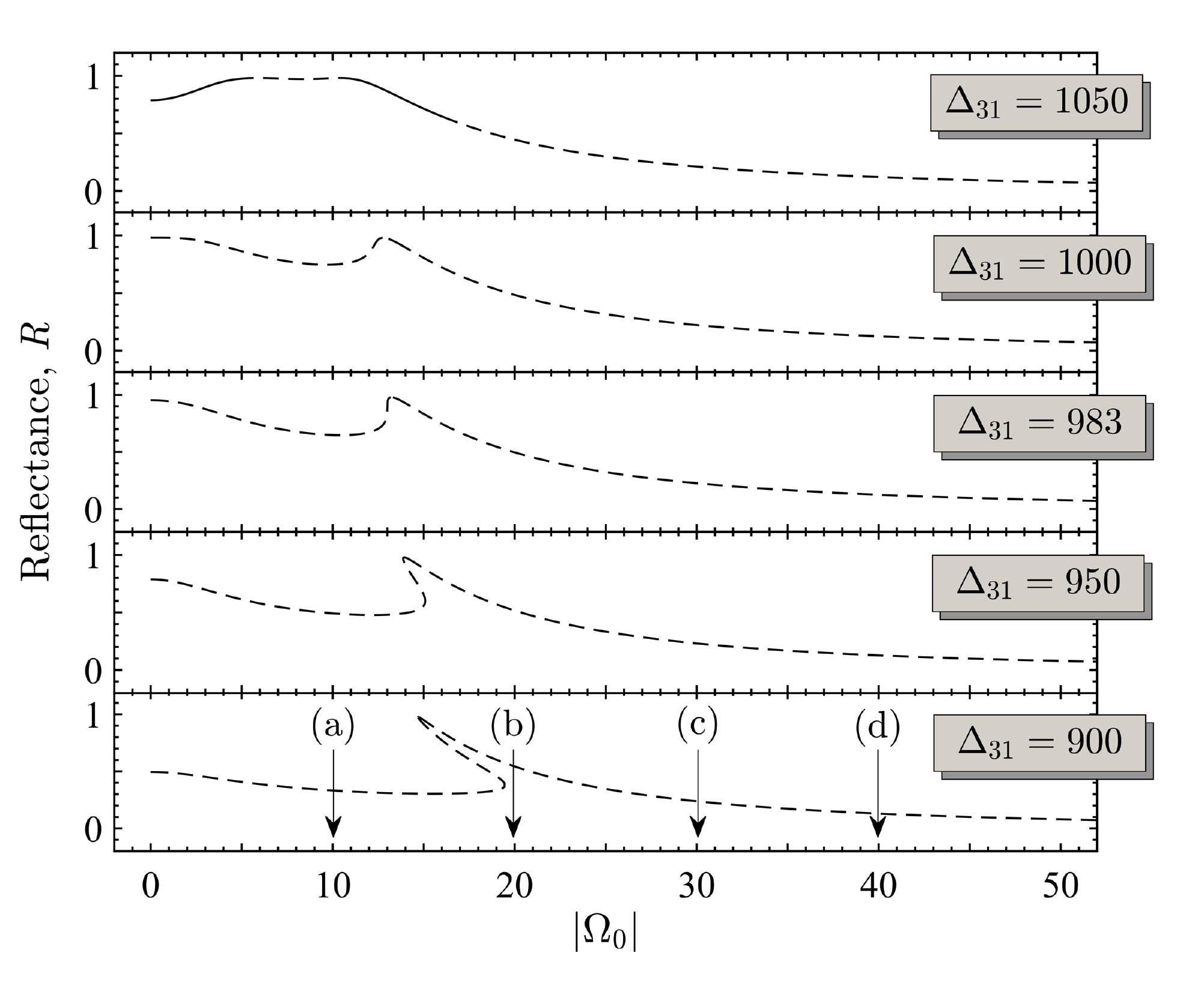}}
\caption{Steady-state reflectance $R$ as a function of the Rabi magnitude $|\Omega_0|$ of the incident field for different values of the detuning $\Delta_{31}$. Other parameters of calculations are: $\Delta_{21} = 100$, $\gamma_{21} = 0.01$. Solid and dashed fragments of the curves show stable and unstable parts of the latter, respectively. Arrows indicate the Rabi magnitudes $|\Omega_0|$ of the incident field for which the reflectance dynamics is calculated, see Fig.~\ref{fig:Reflectance dynamics Delta21 = 100}. All frequency-dependent quantities are given in units of the radiation rate $\gamma_{31}$. }
\label{fig:Steady-state}
\end{figure}

In the nonlinear regime ($|\Omega_0| \gg 1$), both transitions
$1 \leftrightarrow 3$ and $2 \leftrightarrow 3$ contribute to the reflected field.
To solve the nonlinear steady-state problem we made use of the analytical method developed in Ref.~\cite{RyzhovPRA2019}. The typical example of the results, obtained for the doublet splitting $\Delta_{21} = 100$ and the relaxation rate $\gamma_{21} = 0.01$, while varying the detuning $\Delta_{31}$, is presented in Fig.~\ref{fig:Steady-state}.
The solid and dashed fragments of the curves indicate their stable and unstable parts, respectively. To explore the stability of different solutions, we used the standard Lyapunov's exponents analysis~\cite{EckmannRevModPhys1985,NeimarkLandaBook1992,OttBook1993}, calculating the eigenvalues $\Lambda_k$ ($k=1\ldots 8$, eight being the dimensionality of the system's phase space) of the Jacobian matrix of the right hand side of Eqs.~(\ref{rho11})--(\ref{rho21}) as a function of $|\Omega|$~\cite{RyzhovPRA2019}. The Lyapunov's exponent $\Lambda_k$ with the maximal real part $\mathrm{Max}_k\{\mathrm{Re}[\Lambda_k]\}$ determines the character of a given steady-state solution (stable/unstable): if $\mathrm{Max}_k\{\mathrm{Re}[\Lambda_k]\} \leq 0$ the solution is stable and unstable otherwise. Surprisingly, the $R-vs-|\Omega_0|$-dependence for selected values of $\Delta_{31}$ appears to be unstable almost in the whole range of $|\Omega_0|$ considered, including $|\Omega_0|\lesssim 1$. Additionally, the reflectance $R$ may have several solutions (up to three for $\Delta_{31} > 983$) for a given value of $|\Omega_0|$ with $\Delta_{31} = 983$ being the threshold for a thee-valued solution to occur, The multiplicity of solutions implies bistability and hysteresis behavior of reflectance~\cite{RyzhovPRA2019}.

\begin{figure*}[ht!]
\centering
\fbox{\includegraphics[width=0.9\linewidth]{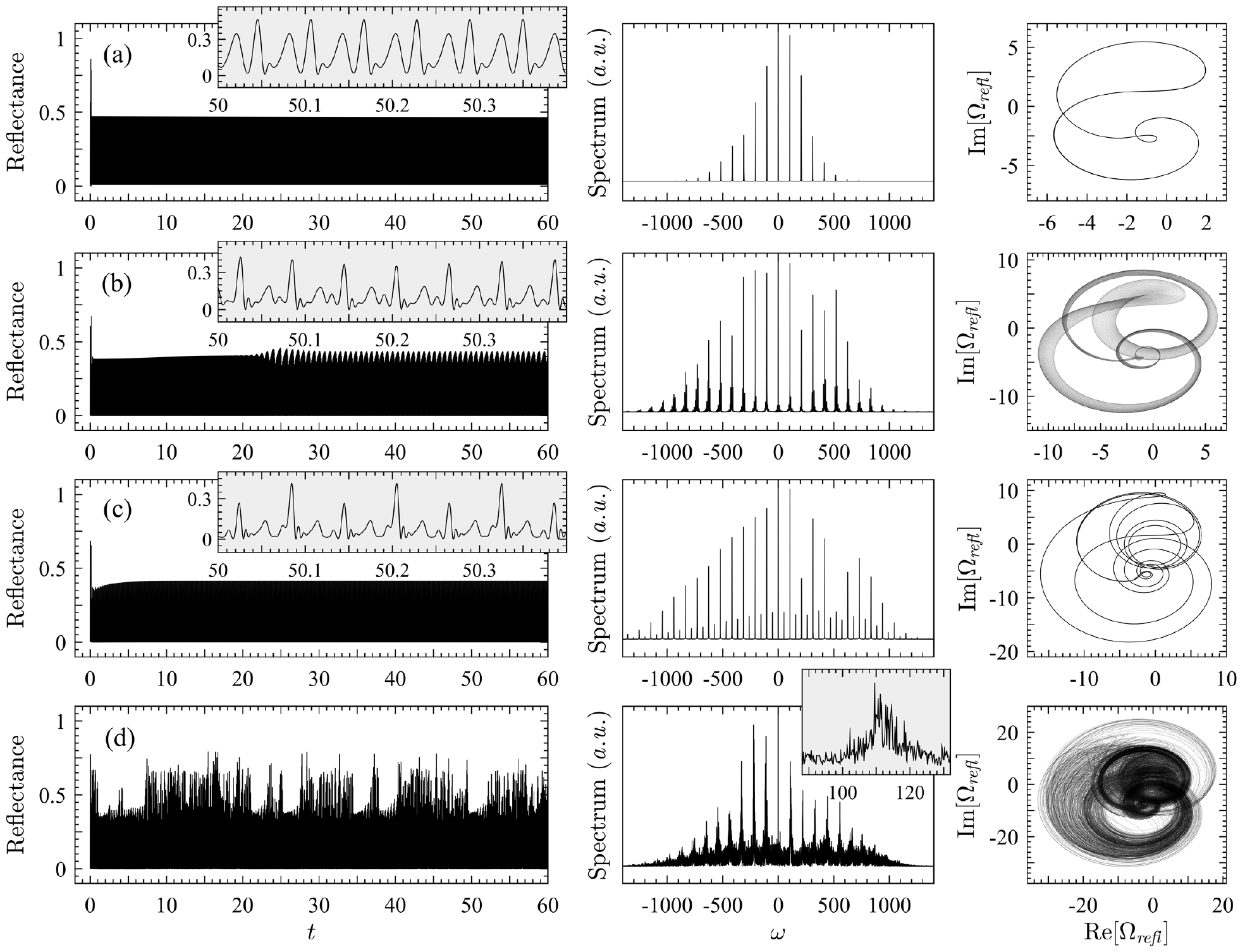}}
\caption{Time-domain behavior of the reflectance $R$ (left panels), the Fourier spectrum, $|\int_{T} \exp{i\omega t} \Omega_\mathrm{refl}(t) dt|$, and the two-dimensional phase-space map ($\mathrm{Re}[\Omega_\mathrm{refl}], \mathrm{Im}[\Omega_\mathrm{refl}]$) of the attractor (right panels) obtained by solving Eqs.~(\ref{rho11})--(\ref{Local field}) for the ground-state initial condition, $\rho_{11}(0) = 1$, and four values of the Rabi magnitude $|\Omega_0|$ of the incident field shown in Fig.~\ref{fig:Steady-state} by arrows. The parameters of calculations are: $\Delta_{31} = 900$, $\Delta_{21} = 100$, $\gamma_{21} = 0.01$. The inserts blow up the details of dynamics. All frequency-dependent quantities are given in units of the radiation rate $\gamma_{31}$, while time is in units of $\gamma_{31}^{-1}$. }
\label{fig:Reflectance dynamics Delta21 = 100}
\end{figure*}
{\it Time-domain}.
To uncover the character of reflectance instabilities, we performed time-domain calculations for several values of the Rabi magnitude $|\Omega_0$ of the incident field, indicated by arrows in Fig.~\ref{fig:Steady-state}, and the ground state initial conditions ($\rho_{11}(0) = 1$ while all other density matrix elements are equal to zero), Equations~(\ref{rho11})--(\ref{rho21}) were integrated until all transients vanish and the system reaches a sustainable phase -- attractor -- which further was analized on an interval $T$. More specifically, we calculated the attractor's Fourier spectrum $|\int_{T} \exp{i\omega t} \Omega_\mathrm{refl}(t) dt|$ and the two-dimensional phase-space map ($\mathrm{Re}[\Omega_\mathrm{refl}], \mathrm{Im}[\Omega_\mathrm{refl}]$). The results are presented in Fig.~\ref{fig:Reflectance dynamics Delta21 = 100}.

Shown in Fig.~\ref{fig:Reflectance dynamics Delta21 = 100} are: left panels -- time-domain behavior of the Rabi magnitude $|\Omega_\mathrm{refl}(t)|$ of the reflected field, the Fourier spectrum (middle panels) and two-dimensional phase-space map (right panels) of the attractor for four values of $|\Omega_0|$ indicated by arrows in Fig.~\ref{fig:Reflectance dynamics Delta21 = 100}, (panel $\Delta_{31} = 900$): (a) - $|\Omega_0| = 10$, (b) - $|\Omega_0| = 20$, (c) - $|\Omega_0| = 30$, and (d) - $|\Omega_0| = 40$.

As observed from Fig.~\ref{fig:Reflectance dynamics Delta21 = 100}, the reflectance dynamics exhibits various types of attractors: for (a) and (c) cases, it evolves towards limit cycles, which has its confirmation in the equidistant character of the attractor's Fourier spectrum and in closeness of the attractor's trajectory. Accordingly, the reflectance dynamics represents periodic self-oscillations. Note that for the set of parameters used, the frequencies of self-oscillations reside in THz domain.

Oppositely, for the case (b), the attractor's Fourier spectrum, in addition to harmonics of the base frequency, contains satellites with incommensurate frequencies, implying an aperiodic motion -- aperiodic self-oscillations. And finally, the case (d) resembles a chaotic behavior of reflectance: the attractor's Fourier spectrum is of a quasi-continuous nature and the trajectory densely covers a finite area in the phase space.

Alternating the character of motion on changing the Rabi magnitude $|\Omega_0|$ of the incident field means that the system undergoes bifurcations~\cite{GuckenheimerBook1986,Arnol'dBook1994}. A detaled study of this phenomenon represents a stand-alone problem and will be addressed elsewhere.

{\bf Conclusion}.
In conclusion, we have conducted a theoretical study of reflectance of a metasurface comprizing regularly spaced quantum $\Lambda$-emitters subjected to a CW quasi-resonant excitation. We have found that in the vicinity of the collective (excitonic) resonance the monolayer almost totally reflects the incident field, thus acting as a nanometer-thin resonant mirror. Moreover, within a certain range if frequencies, the reflectance turns out to be a three-valued function of the incident field magnitude, implying bistability and hysteresis behavior.

Using the Lyapunov's exponent analysis, we have found windows of stability and instability of the reflectance-versus-incident field magnitude dependence and  unraveled their character by solving the time-domain problem. It has turned out that, depending on the incident field magnitude, the system may exhibit a variety of instabilities, such as periodic and aperiodic self-oscillations, and chaotic behavior. The (secondary) field, acting on an emitter on the part of the others, provides a positive feedback which gives rise to instabilities found.

Our results suggest various practical applications of metasurfases of quantum $\Lambda$-emitters, such as a nanometer-thin bistable mirror, a tunable generator of coherent THz radiation (in self-oscillation regime), and an optical noise generator (in chaotic regime), which makes the considered system promising for nanophotonics.

R. F. M. acknowledges M. Akmullah Bashkir State Pedagogical University for a financial support.


\end{document}